\documentclass{emulateapj}

\usepackage{apjfonts}
\usepackage{amsmath}
\usepackage{amssymb}
\usepackage{natbib}
\usepackage{xspace}
\usepackage{graphicx}
\usepackage{rotate} 
\usepackage{subfigure}
\usepackage{float}

\newcommand{\err}[2]{\ensuremath{^{+#1}_{-#2}}\xspace}
\newcommand{\e}[1]{\ensuremath{\times 10^{#1}}\xspace}

\newcommand{\Msun}{\ensuremath{M_\odot}\xspace}

\newcommand{\nhgal}{$N_{\text{H, Gal}}$\xspace}
\newcommand{\etal}{et al.~}
\newcommand{\ka}{K$\alpha$\xspace}
\newcommand{\kb}{K$\beta$\xspace}

\newcommand{\fluxunits}{erg\,cm$^{-2}$\,s$^{-1}$\xspace}
\newcommand{\plnormunits}{ph\,keV$^{-1}$\,cm$^{-2}$\,s$^{-1}$ at 1 keV\xspace}
\newcommand{\chidof}{$\chi^{2}/$dof\xspace}
\newcommand{\dchidof}{$\Delta \chi^{2}$/dof}

\newcommand{\xte}{\textsl{RXTE}\xspace}

\newcommand{\suzaku}{\textsl{Suzaku}\xspace}

\newcommand{\xmm}{\textsl{XMM-Newton}\xspace}
\newcommand{\chandra}{\textsl{Chandra}\xspace}
\newcommand{\pn}{EPIC-pn\xspace}

\newcommand{\mkn}{Mkn~590\xspace}
\newcommand{\Long}{Longinotti \etal (2007)\xspace}
\newcommand{\xspec}{\textsc{xspec}\xspace}

\newcommand{\pexrav}{\textsc{pexrav}\xspace}

\shorttitle{Mkn 590}
\shortauthors{Rivers et al.}

\begin{document}

\title{A Suzaku Observation of Mkn 590 Reveals a Vanishing Soft Excess} 

\author{Elizabeth~Rivers\altaffilmark{1}, Alex~Markowitz\altaffilmark{1,2,3}, Refiz~Duro\altaffilmark{2}, Richard~Rothschild\altaffilmark{1}}
\affiliation{1 University of California, San Diego, Center for Astrophysics and Space Sciences, 9500 Gilman Dr., La Jolla, CA 92093-0424, USA\\
2 Dr.\ Karl Remeis-Sternwarte and Erlangen Centre for Astroparticle Physics, Frederic-Alexander Universit\"{a}t Erlangen-N\"{u}rnberg, 
 7 Sternwartstrasse, 96049 Bamberg, Germany\\
3 Alexander von Humboldt Fellow}


\begin{abstract}

We have analyzed a long-look \suzaku observation of the Seyfert 1.2 Mkn~590.  We aimed to measure the Compton reflection
strength, Fe K complex properties and soft excess emission as had been observed previously in this source.  
The Compton reflection strength was measured to be in the range 0.2--1.0 depending on the model used.
A moderately strong Fe \ka emission line was detected with an equivalent width of $\sim$120$\pm25$ eV and an Fe \kb line
was identified with an equivalent width of $\sim$30$\pm20$ eV, although we could not rule out contribution from ionized Fe
emission at this energy.  Surprisingly, we found no evidence for soft excess emission.  
Comparing our results with a 2004 observation from \xmm we found that either the soft excess 
has decreased by a factor of 20--30 in 7 years or the photon index has steepened by 0.10 (with no soft excess present) 
while the continuum flux in the range 2--10 keV has varied only minimally (10\%).  
This result could support recent claims that the soft excess is independent of the X-ray continuum.

\end{abstract}

\keywords{galaxies: active  -- galaxies: individual (\mkn) -- X-rays: galaxies}


\section{Introduction}
The physical conditions and geometry of the accreting circumnuclear material in the vicinity of supermassive black holes in active galactic nuclei (AGNs) 
remain open and critical questions. Studying AGN X-ray spectra and spectral variability yields constraints on the geometry of circumnuclear accreting material.
Studying AGN X-ray spectra and spectral variability yields constraints on the geometry of circumnuclear accreting material.  
The X-ray continuum for Seyferts can be modeled as a power law arising in a hot corona via inverse Compton scattering of soft seed photons
(e.g., Haardt et al.\ 1994) and can display rapid variability.
This continuum may be reprocessed by material in the vicinity of the black hole, contributing emission lines and 
the Compton reflection hump which is commonly seen in Seyfert spectra above $\sim$10 keV (Nandra \& Pounds 1994; Rivers \etal 2011).  
The Fe K$\alpha$ emission line is nearly ubiquitous in AGN spectra and is a key diagnostic as its width can indicate the distance 
from the supermassive black hole of the emitting material and its energy can indicate an origin in neutral or ionized material.

Excess emission above the continuum at energies less than $\sim$2 keV, known as the soft excess, 
is commonly seen in Seyferts (e.g., Arnaud \etal 1985), yet remains something of a mystery. 
While a phenomenological blackbody component has generally been found to be a good fit in the past (e.g.,  Piconcelli \etal 2005),
the invariance of the inferred temperature across a wide range of luminosities and black hole masses
has led to the general consensus that the soft excess is non-thermal in origin.
Reflection from an ionized medium (Ross \& Fabian 2005's ``{\tt REFLION}'' or the updated  ``{\tt REFLIONX}'' ) 
offers a blend of ionized soft emission lines to form the soft excess; blurring of such emission has been seen to 
fit many Seyfert spectra well (Crummy \etal 2006). 
The presence of Fe XXV and/or Fe XXVI emission lines can also indicate the presence of ionized material, but these are 
relatively rare and have been reported for only about a dozen Seyferts (Bianchi et al.\ 2005).

Mehdipour \etal (2011) observed Mkn 509 several times over a period of $\sim$100 days with \xmm and discovered that variations in
the soft excess were not correlated with variations in the 2--10 X-ray power law, but were correlated with variations
in the UV flux.  This suggests that the soft excess does not arise from reflection processes off ionized material (see, e.g., Ross \& Fabian 2005).
Instead, one possibility is that the soft excess is created through thermal Comptonization of optical/UV photons from the disk at a location separate from
the hard X-ray continuum with a lower temperature and/or optical depth causing a softer spectrum.

The Seyfert 1.2 galaxy Mkn 590 (also known as NGC 863) is an excellent candidate to explore the nature of circumnuclear material in AGNs.  
It is a relatively X-ray bright, reverberation-mapped object with a well determined black hole mass ($M_{\rm BH} = 4.90^{+0.97}_{-0.99}
\times 10^{7}\,\Msun$; Vestergaard \& Peterson 2006).  It was observed in 2004 by \textit{XMM-Newton} and \textit{Chandra}
(Longinotti \etal 2007), showing narrow Fe K emission lines from neutral Fe, and possibly from He-like Fe and H-like Fe.
Those authors modeled a Compton reflection hump, which, along with the strong neutral Fe line and lack of a relativistic line broadening,
suggested emission from a distant cold torus.  However, lacking coverage above 10 keV, they  
could provide only weak constraints on the total Compton reflection strength, $R$ ($R=1$ corresponds to reflecting material 
covering 2$\pi$\,sr as seen from the illuminating source; Magdziarz \& Zdziarski 1995).
A \textit{Rossi X-Ray Timing Explorer} (\xte) spectrum from public archive data contains only $\sim$30 ks of good exposure time, 
also yielding extremely poor constraints on $R$ (Rivers \etal in prep).  
\textit{BeppoSAX} did not observe this source.  

Longinotti et al.\ (2007) ruled out reflection from an ionized disk as the source of the soft excess,
because it did not fit the ionized emission lines well.  They concluded that the Fe XXV and XXVI emission lines 
originated in photo-ionized, optically-thin gas in an extended medium.  Such emission lines are commonly seen 
in the X-ray spectra of Seyferts 2's, but not Seyfert 1-1.5's, however they are usually limited to emission lines at soft 
X-rays, not ionized Fe K lines.  Importantly, Longinotti et al.\ (2007) found no evidence for an ionized absorber lying 
along the line of sight, which can adversely affect modeling of continuum emission components 
(e.g., Reeves et al.\ 2004, Turner et al.\ 2005).  The only soft X-ray emission line was from O VIII, which they also attributed to
photo-ionized circumnuclear gas. In other words, Mkn 590 has a relatively clean line of sight to the nucleus and can be considered a
``bare nucleus'' like the Seyfert 1's Ark 120 (Vaughan et al.\ 2004) or MCG--2-58-22 (Weaver et al.\ 1995).

In this paper we will analyze and discuss results from an observation by the \suzaku observatory
performed in 2011 January.  Our goals with this observation were to take advantage of \suzaku's
broadband coverage to better constrain the form of the hard X-ray
spectrum, including the hard X-ray behavior above 10 keV, and to search for long-term spectral
variability between the 2004 and 2011 observations.  In Section 2 we will detail the data reduction process;
in Section 3 we will describe our spectral fitting methods; and in Section 4 we will discuss our results.

\section{Data Reduction}

\suzaku observed Mkn 590 on 2011 January 23 but was interrupted by a TOO trigger.
The observation was continued on 2011 January 26.
Data were taken with \suzaku's X-ray Imaging Spectrometer (XIS; Koyama \etal 2007) and 
Hard X-ray Detector (HXD; Takahashi \etal 2007), and processed using version 2.5.16.29 of the \textsl{Suzaku} pipeline 
with the recommended screening criteria (as per the \textsl{Suzaku} Data Reduction 
Guide\footnote{http://heasarc.gsfc.nasa.gov/docs/suzaku/analysis/abc/abc.html}).  
All extractions were done using HEASOFT v.6.9.


\subsection{XIS Reduction}

The XIS is comprised of one back-illuminated (BI) and two front-illuminated (FI) CCD cameras\footnote{The fourth CCD camera has been 
inoperative since 2006 November.  See the \textsl{Suzaku} ABC Guide for details.} each placed in the focal plan of an X-ray Telescope module.
The screened XIS events data were cleaned and the modes summed to create image files for each XIS and from these we extracted source 
and background lightcurves and spectra using a standard extraction radius of 2' for the source and four background regions of equivalent area.  
We then used XISRMFGEN and XISSIMARFGEN to create the response matrix (RMF) and ancillary response (ARF) files. 
Data for the two observation windows were added after we confirmed that there was no significant change in 
flux or spectral shape between the two observations.  The total good exposure time per XIS was 102 ks.
We also co-added data from the two FI CCDs once we had confirmed that the spectra were consistent with each other.

\begin{figure}
  \plotone{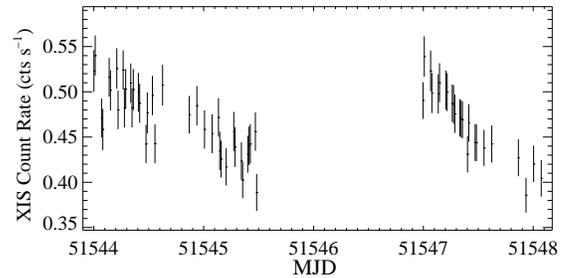}
  \caption{Background subtracted 2--10 XIS light curve for the 2011 January 23--26 \suzaku observation averaged between all three working XIS CCD's.}
  \label{figlc}
\end{figure}

Data were grouped with a minimum of at least 50 counts per bin.
We ignored data above 12 keV (10 keV for BI) where the effective area of the XIS begins to drop dramatically and 
below 0.7 keV due to time-dependent calibration issues of the instrumental O K edge at 0.5 keV.
We included additional Gaussians in the fitting procedures between 1.5 and 2.4 keV due to calibration uncertainties 
for the Si K complex and Au M edge arising from the detector and mirror system. 
Average 2--10 keV rates were 0.442$\pm 0.002$ and 0.538$\pm 0.002$ counts\,s$^{-1}$ per XIS for FI and BI, respectively.
Figure \ref{figlc} shows the average XIS light curve for the duration of the observation. 

\subsection{PIN Reduction}

The HXD consists of two detectors, the PIN diodes and the GSO scintillators, 
however we did not analyze the GSO data due to the faintness of 
the source relative to the non-X-ray background in the GSO band.  
The PIN is a non-imaging instrument and the HXD instrument team provides non-X-ray background model event files for each observation.
Instrument background estimates yield $\lesssim$\,1.5\% systematic uncertainty at the 1$\sigma$\ 
level (Fukazawa \etal 2009).  We simulated the Cosmic X-ray Background in \xspec ver.12.6 using the form of 
Boldt (1987) as recommended in the \suzaku ABC Guide.

We extracted net spectra which were then deadtime corrected for a net exposure time of 86~ks.  
As with the XIS, data were grouped manually but with a minimum of at least 200 counts per bin.
We excluded PIN data below 15 keV due to thermal noise and above 45 keV where the source 
became too faint to detect above the background.


\section{Spectral Fitting}

All spectral fitting was done in \xspec version 12.7.1, utilizing solar abundances of Wilms \etal (2000) and cross-sections from Verner \etal (1996). 
Uncertainties are listed at the 90\% confidence level ($\Delta \chi^2$ = 2.71 for one interesting parameter).

We began by fitting the broad spectrum (XIS+PIN) from 0.7 keV to 45 keV with a power law absorbed by a Galactic column of 
\nhgal=2.65$\times 10^{20}$\,cm$^{-2}$ (Kalberla \etal 2005).  We included a fixed constant factor of 1.16 between the XIS-FI and PIN
data as well as a free constant between the FI and BI data (expected to be very close to 1).  We left the photon index free between the FI and
BI due to instrumental flattening in the BI, however in all cases the values were consistent with each other.
We also included three narrow Gaussians to account for residuals associated with the instrumental Si K complex and Au M edge.
For the XIS-FI data a Gaussian with a negative normalization at 1.85 keV and a positive Gaussian at 2.26 keV accounted for these residuals.
For the XIS-BI data three positive Gaussians at 1.49, 1.79 and 2.26 keV accounted for the residuals (see, e.g., Suchy \etal 2011).
These energies were fixed for our fitting.

Residuals at 6.4 keV clearly showed the need for a neutral Fe line and broad residuals in the PIN data indicated the presence of Compton 
reflection (Figure 2b).  We added a Gaussian to model the Fe \ka line and a \pexrav component to model reflection with the photon index and
normalization tied to that of the incident power law with the reflection strength, $R$, as the only free parameter (cos$i$ fixed at 0.45 and all 
abundances set to solar).  Note that the systematic uncertainty in the PIN background did not contribute significantly to the total uncertainty on $R$ 
($\Delta R_{\rm Sys}$ < 0.01).  The addition of these components yielded a good fit with \chidof=198/218.  Parameters for this model are listed in 
Table \ref{tabpar} as our ``basic'' model.  We measured a 2--10 keV flux of 6.8 \e{-12} \fluxunits, corresponding to a luminosity of 
$L_{2-10} = 8\times 10^{42}$ erg s$^{-1}$  assuming a luminosity distance of 105 Mpc.

We found no evidence for the presence of absorption in excess of the Galactic column nor a soft excess 
below 2 keV.  The latter is a surprising result since Longinotti \etal (2007) found clear evidence of a soft excess in their 2007 \xmm data.  They 
found \dchidof=$-$49/2 for a blackbody with $kT$=156\err{14}{12} eV and \dchidof=$-$53/3 for a soft X-ray power law with $\Gamma$=1.89$\pm0.03$.  
Adding a soft excess component to our model did not improve the fit and yielded an upper limit to the normalization of a soft power law of 
3\e{-5} \plnormunits ($\Gamma \sim 2.5$) or using the blackbody model {\sc BBODY}, an upper limit to the normalization of
$3\times10^{-7}$ (with $kT$ fixed at 156 keV) corresponding to a blackbody luminosity of $L \leq 3\times 10^{40}$ erg s$^{-1}$ (see \xspec user's guide
for details on the {\sc BBODY} normalization).


\begin{deluxetable*}{l|c|cc|cc}
   \tablecaption{Parameters for Best-fit Models \label{tabpar}}
   \tablecolumns{6}
   \startdata
\hline
\hline\\[-0.4mm]
Model  &   Basic    &     \multicolumn{2}{c}{Basic ($\Gamma$ Untied)}   &  \multicolumn{2}{c}{Basic + Soft Excess} \\[1mm] 
Data  &    XIS+PIN    &   XIS+PIN & \pn  &   XIS+PIN & \pn  \\[1mm]
\hline \\
$\Gamma$    		&	1.67$\pm0.01$		&	1.71$\pm0.01$	&	1.82$\pm0.01$	&	1.67$\pm0.01$		& -	\\[0.5mm]
$A_{\rm PL}\tablenotemark{a}(10^{-4})$&	16.6$\pm0.1$		&	16.4$\pm0.1$	&	-			&	16.4$\pm0.1$ 		& -	\\[0.5mm]
$\Gamma_{SXPL}$  &					&				&				&	  3.0$\pm0.3$	  & - 	\\[0.5mm]
$A_{SXPL}\tablenotemark{a} (10^{-4})$ &					&				&				&	< 0.1 		& 2.9$\pm0.5$	\\[0.5mm]
Fe \ka Line E (keV)	&	6.43$\pm0.02$		&	6.43$\pm0.02$ &  	-		&	6.43$\pm0.02$ 	& -  \\[0.5mm]
................. $I_{\rm Fe}\tablenotemark{b} (10^{-5})$ & 1.0$\pm0.2$	&	0.8$\pm0.2$	&	0.9$\pm0.2$	&	1.0$\pm0.2$ 	& 1.2$\pm0.2$ \\[0.5mm]
................. $\sigma$ (eV) &	< 80				&	< 60			&	-			&	< 60 			& -		\\[0.5mm]
................. EW (eV)	&	120$\pm25$		&	90$\pm20$	&	120$\pm30$	&	110$\pm15$ 	& 130$\pm20$	\\[0.5mm]
Fe \kb Line E (keV)     &   	7.0\err{0.1}{0.4} 	   	&	7.03$\pm0.04$	&		&	7.02$\pm0.04$				& 	\\[0.5mm]
................. $I_{\rm Fe}\tablenotemark{b} (10^{-5})$ & 0.25$\pm0.15$  &	< 0.2	&	0.4$\pm0.1$	&	 0.17$\pm0.15$	&  0.5$\pm0.2$   \\[0.5mm]
................. EW (eV)	&	30$\pm20$		&	< 20	&	60$\pm15$	&	25$\pm20$ 	& 70$\pm30$	\\[0.5mm]
$R$				&	0.3$\pm0.2$		&	1.3$\pm0.2$ 	&  	-			&	0.8$\pm0.2$ & -		\\[0.5mm]
\chidof			&	199/218			&	550/437		&	-			&	478/435		& -	\\
\enddata
\tablecomments{Parameters from our best-fit models.  The first column gives parameters to our best fit model of \suzaku data only consisting of a power law with Galactic absorption, an Fe \ka line and Compton reflection.  The next column gives the results of fitting \suzaku and \xmm data simultaneously with the basic model from the first column with $\Gamma$ and overall normalization untied between the two observations.  The next column shows the same data set but with $\Gamma$ tied and a power law soft excess component with normalization untied between the two observations.
A dash indicates a tied parameter.}
\tablenotetext{a}{Power law normalization (ph\,keV$^{-1}$\,cm$^{-2}$\,s$^{-1}$ at 1 keV)}
\tablenotetext{b}{Fe \ka line intensity (ph\,cm$^{-2}$\,s$^{-1}$)}
\end{deluxetable*}

\subsection{Simultaneous Fitting of \suzaku and \xmm Data}

Searching for the source of this discrepancy we re-reduced the 2007 \xmm data and analyzed the \pn spectrum between 0.6 and 10 keV.  
The EPIC-pn camera (Str\"uder \etal 2001) was operated in small window mode with medium filter in place in order to prevent possible photon pile-up.
The data were reduced with version 11.0.0 of the XMM-Newton Science Analysis Software (SAS),
following the standard data reduction steps.  The good exposure time after screening was 69 ks for the EPIC-pn. 
The radius of the source area was chosen to be 40 arcseconds 
and the background was extracted from the areas with no source photon contamination. 
We used the SAS tool \texttt{epatplot} to check for the presence of  photon pile-up, finding none.
While we have used more recent calibration files, model fitting resulted in values consistent with those of Longinotti \etal (2007).
We found that the 2--10 keV flux of the source in 2004 was very similar to our 2011 observation (7.5 $\times 10^{-12}$ erg\,cm$^{-2}$\,s$^{-1}$).  
We therefore decided to try fitting the \xmm and \suzaku data simultaneously.  
The combined data set and residuals to the models we applied are shown in Figure \ref{figspecpn}.

We applied our best fit ``basic'' model derived from the \suzaku spectrum (power law with Galactic absorption plus Compton hump and Fe lines,
with no soft excess) to both spectra simultaneously, allowing for a free constant renormalization factor between the XIS and EPIC-pn,
but otherwise keeping all spectral parameters (including $\Gamma$) tied.  It was not significant to leave the Fe line normalization untied.
This yielded a poor fit with \chidof= 782/438 and confirmed that a spectral change had indeed taken place between the two observations.
Residuals for this fit are shown in Figure \ref{figspecpn}b and a clear divergence can be seen between the two observations.
We next untied the photon index between the two observations.  This yielded an acceptable fit with  \chidof = 550/437. 
We also fit a model with the photon index tied as before, but including an additional power law to model the soft excess with its normalization 
free between the two observations.  This yielded an improved fit with  \chidof = 478/435.  
Parameters for both of these fits are listed in Table \ref{tabpar}.  Results using a blackbody soft excess model were qualitatively similar 
with the blackbody temperature tied between the data sets yielding a best fit $kT$ of 180$\pm10$ eV, a normalization for the EPIC-pn data 
of (7.9$\pm0.7$)\e{-6} corresponding to a luminosity of (8$\pm1$)\e{41} erg s$^{-1}$ and for \suzaku an upper limit of 3\e{-7} or blackbody 
luminosity of $L \leq 3\times 10^{40}$ erg s$^{-1}$, as expected from previous fits.

\subsection{Fe K Complex Analysis}

Next we focused on the mid-X-ray range (4--10 keV) in order to examine the Fe complex in more detail, using only the XIS-FI data since it 
has the best sensitivity in this range.  We began with the model from the broad fitting, freezing the continuum and Compton reflection parameters 
since these parameters are best determined from the broad fit.   Note that the Fe edge at 7.11 keV is included in the \pexrav model and that the depth
was therefore a fixed quantity in these fits.  We noticed residuals around 7 keV which could indicate the presence of an Fe XXVI emission line arising in 
highly ionized material or an Fe \kb line arising in mostly neutral material.  We added a Gaussian component with $\sigma$ tied to that of the 
Fe \ka line and found a best fit energy centroid of 7.03\err{0.05}{0.12} keV and an intensity of $\leq$3.6\e{-6} ph\,cm$^{-2}\,$\,s$^{-1}$ 
or $\sim$\,40\% of the Fe \ka line.  The improvement in fit when adding this component was \dchidof = $-$6/3, which is not a substantial 
improvement, however it is consistent with the presence of either Fe XXVI or a \kb line.  Adding an Fe \kb line with energy frozen at 
7.056 keV and intensity equal to 13\% that of the Fe \ka line resulted in an upper limit to a Fe XXVI of 2.8$\times 10^{-6}$ ph\,cm$^{-2}$\,s$^{-1}$.
Testing for the additional presence of Fe XXV emission did not reveal a significant detection; the upper limit to the line intensity was
9\e{-6} ph\,cm$^{-2}$\,s$^{-1}$.  We tested for an additional broad Fe \ka line but found only an upper limit for the normalization of 
$4.4\times 10^{-6}$ ph\,cm$^{-2}$\,s$^{-1}$ and an $EW \lesssim 70$ eV using the \textsc{diskline} model in \xspec with $R_\text{in}$, and 
$\beta$ as free parameters (however their values were poorly constrained by the fit since the normalization of the line was consistent with 0).


\begin{figure}
  \plotone{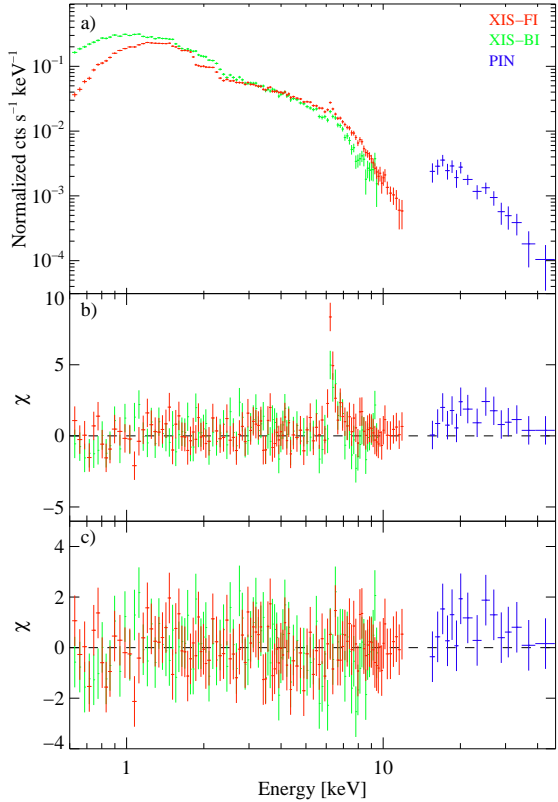}
  \caption{\suzaku XIS and PIN data for the January 2011 observation.  Panel a) shows the data; b) residuals to a simple power law plus Galactic 
  absorption; c) residuals to the best fit model with Fe \ka and Compton reflection modeled (but no soft excess component).}
  \label{figspec}
\end{figure}


\begin{figure}
  \plotone{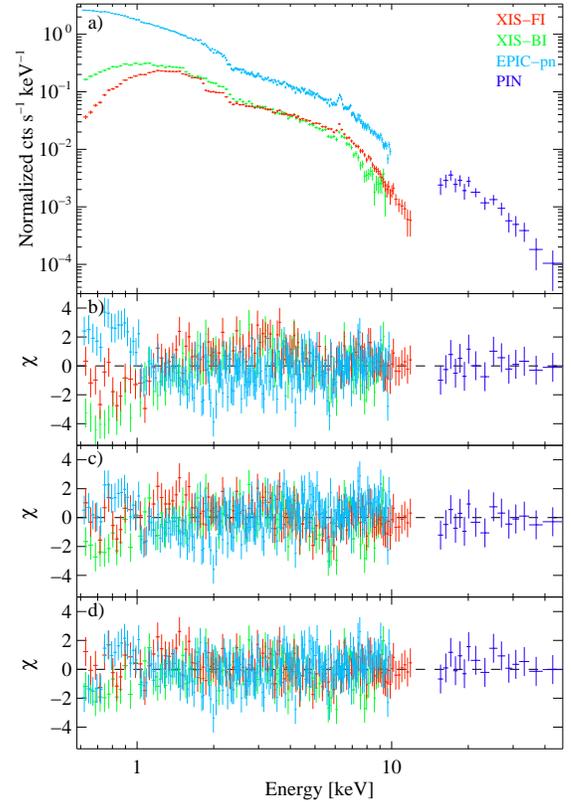}
  \caption{Fitting \suzaku and \xmm data together.  Panel a) shows data from the \suzaku XIS and PIN and from the EPIC-pn; b) shows residuals
  to our basic fit with no soft excess modeled and $\Gamma$ tied between the two data sets; c) shows residuals to the basic model with $\Gamma$
  untied; d) shows residuals to the basic model plus a soft excess power law component (normalization untied).}
  \label{figspecpn}
\end{figure}

\section{Discussion and Conclusions}

The \suzaku spectrum of Mkn~590 shows many components typical to Seyferts, including a moderate photon index and a 
strong narrow Fe \ka line.  We found an upper limit to the width of the Fe \ka line of $\sigma <$ 80 eV corresponding to a 
FWHM velocity of $v_{\rm FWHM} \,\leq\, 8900$ km s$^{-1}$, assuming $\frac{3}{4} v_{\rm FWHM}^2= \langle v_{\rm disp}^2 \rangle$ (Netzer 1990). 
Assuming Keplerian motion this corresponds to a distance of $\gtrsim$0.0026 pc or 3.1 lt days.  
This is consistent with results from Longinotti \etal (2007) who found a resolved \ka line with a FWHM velocity 
of 4000\err{2000}{2700} km s$^{-1}$.  Both of these results are consistent with measurements of the H$\alpha$ line width of 
2000 km s$^{-1}$ (Stirpe 1990), however it remains unclear if the Fe emission is associated with the broad line region.
We also measured a line at $\sim$7.0 keV which is likely Fe \kb emission from neutral or at most moderately ionized Fe, 
however we cannot rule out a contribution from Fe XXVI.

We measured the strength of the Compton reflection component in this source, however the amount of reflection we measure is highly
dependent on the model chosen.  In the best fit to the \suzaku data we found a relatively low value for $R$, however residuals
for the PIN data are consistently above the line in this fit.  This could indicate that there is a stronger reflection component, 
but increasing the value manually leads to significant residuals between 5 and 10 keV in the XIS, particularly for the BI.
Allowing the PIN normalization to be free reduced the value of $R$ to $\leq\,$0.1 but with a normalization constant of 1.65 between
the PIN and the XIS, much higher than calibration by the instrument team allows it should be.
We conclude that it is likely the Compton reflection strength in this source is in the range $R$=0.2--1 based on the basic fit to the 
\suzaku data and the soft excess model fit to the combined data.
A more sensitive hard X-ray detector or a longer integration time would be required to give more definitive results.

\subsection{The Soft X-ray Conundrum}

The flux of the source measured in 2011 January  was very similar to the \xmm measurement from 2004.
Given the similarity in flux level for these two observations, we did not expect significant deviation in the spectral characteristics
of the source.  However it was clear that no soft excess was detected in the \suzaku data.  If indeed a soft excess was detected in the
\xmm observation then it has weakened by at least a factor of $\sim$30 in 7 years, while the 2--10 keV flux has remained relatively constant.  
If instead the \xmm observation captured the source in a softer state, then $\Gamma$ has hardened by $\sim$\,0.10, a significant 
change given that the source is at a virtually identical flux level.  
Sobolewska \& Papadakis (2009) analyzed nine Seyferts with longterm monitoring data from \xte, and measured flux, observed $\Gamma$ and in most cases Fe line EW.  Their Figure 8 shows that on average a change in $\Gamma$ of 0.1 would be accompanied by a change in the 2--10 keV flux by a factor of 1.5--2.
This is not to say that this source is not variable.  Mkn 590 was monitored over the course of $\sim$1 year between 2000 and 2001 
by \xte which observed a range of 2--10 keV fluxes of 1--5 \e{-11} \fluxunits (Markowitz \& Edelson 2004).  
{\it Swift}-BAT measured an average flux in the 14--195 keV range of 3.7 \e{-11} \fluxunits corresponding to  
$F_{2-10}\sim$8.6 \e{-12} \fluxunits (Tueller \etal 2008), similar to both the \suzaku and \xmm observations.

We have looked into the possibility of contamination by sources in the field of view such as from an ultra-luminous X-ray source (ULX) or background 
AGN.  \Long found two serendipitous sources in the \xmm field of view but neither was close enough (each around 190'') to contaminate 
either the \xmm or \suzaku data.  The 2004 \chandra ACIS data revealed a third source only 7'' offset from Mkn 590, however the flux of this source
was only 3 \e{-14} \fluxunits in the 0.3--10 keV band or roughly 1.4 \e{-14} \fluxunits in the 0.3--2 keV range.  \Long noted that the source is at the same 
luminosity distance as Mkn 590 of 105 Mpc so that this source would have a luminosity of $L_{0.3-10} \sim\,$4.3\e{40} erg s$^{-1}$, consistent with a 
ULX at a distance of $\sim$3.6 kpc from the AGN.  This is two orders of magnitude 
lower than the flux of the AGN and about fifty times lower than that of the soft component detected in the \xmm data in the 0.3--2 keV range.
It therefore seems most likely that the variability we have observed is intrinsic to the AGN and is simply an unexplained facet of the soft excess.

A weakening or strengthening of the soft excess has been seen in other AGNs, however in most cases it also involved a change in flux state.
For example, Markowitz \& Reeves (2009) reported that NGC 4593 had a soft excess that weakened by a factor of 20 
between an \xmm~ observation and a \suzaku~ observation 5 years later which caught the source in a low hard X-ray flux state.  
However, it has been observed that variations in the soft excess can be independent of those in the hard power-law (e.g., Edelson \etal 2002),
suggesting the possibility of independent physical mechanisms at work.  We can rule out blurred reflection from an ionized medium 
(such as the {\sc REFLION} model from Ross \& Fabian 2005) as a source of the soft excess in this case since our \suzaku data clearly 
show reflection but no soft excess.  A simple blackbody soft excess component provides a good fit to the \xmm data, however this model 
has been shown to clash with theoretical predictions for disk temperatures  (e.g. Gierli\'nski \& Done 2006).

A recent \xmm~ monitoring campaign of Mkn 509 by Mehdipour \etal (2011) revealed a variable soft excess
that correlated with the optical-UV flux but not with the flux of the 2--10 keV power law flux.  Mehdipour \etal (2011) were able to 
successfully apply models describing the soft excess as being produced by thermal Comptonization by a warm (0.2 keV) optically 
thick ($\tau \sim 17$) corona surrounding the inner regions of the disk. This is an important result that may reveal a long sought-after 
link between the X-ray spectral properties of a source and the optical/UV thermal accretion disk emission and could explain the extreme
behavior in Mkn~590.  If a soft excess was present in the past and has recently vanished, then we would expect to see a significant
drop in the UV flux of this source if this link exists.  Unfortunately, there was no optical/UV observation simultaneous with the \suzaku observation.
Future monitoring of this source with simultaneous UV and X-ray instruments would be required to confirm or exclude this hypothesis.

Further monitoring would also add to the small but growing number of observations tracking variations
in the soft excess that are independent of variations displayed by the hard X-ray coronal power law.  
Mehdipour \etal (2011) found a variability range of 1--1.7\e{44} erg s$^{-1}$ for Mkn 509 over the course of 36 days.
Mkn 590 and NGC 4593 have both shown much larger changes, factors of 20--50, on a scale of years.
Turner \etal (2001) reported a decrease by a factor of nearly 3 in the soft excess of the NLSy1 Ark~564 across a 35-day
\textit{ASCA} campaign.  During this time the hard power-law varied strongly on timescales of hours but on average varied 
by a factor of only 1.7 across the whole campaign.  Edelson \etal (2002) noted a qualitatively similar trend for the NLSy1 Ton S180 
with systematic trends on timescales of days--weeks in the soft band independent of variations exhibited above 2 keV.
It should be noted that since both Ark 564 and Ton S180 are narrow line Seyferts they are likely in a different luminosity regime (relative to Eddington)
and may be experiencing different physical processes, particularly with regards to the Compton up-scattered component 
of the soft excess (Done \etal 2012) and direct comparison with these objects should be done with caution.
Nevertheless, understanding how the soft excess varies over a range of timescales could help pin down the source of this feature.
Further investigations which calculate the variability amplitude on multiple timescales would be greatly beneficial,
but would require sustained spectral monitoring with instruments sensitive enough in the soft band to give precise measurements
of the soft excess luminosity.

\begin{acknowledgments}
This research has made use of data obtained from the \textsl{RXTE} satellite, a NASA space mission.
This work has made use of HEASARC online services, supported by NASA/GSFC, and the NASA/IPAC Extragalactic Database, 
operated by JPL/California Institute of Technology under contract with NASA.  
The authors thank J\"orn Wilms and Katja Pottschmidt for valuable input and advice.
The research was supported by NASA Grant NNX11AD07G 
and in part by the European Commission under contract ITN215212 ``Black Hole Universe.''
\end{acknowledgments}


\end{document}